\documentclass[aps,prx,reprint,superscriptaddress,longbibliography,floatfix]{revtex4-2}

\usepackage{graphicx}
\usepackage{amsmath}
\usepackage{amssymb}
\usepackage{xcolor}
\usepackage{listings}
\usepackage{upquote}
\usepackage{tikz}
\usetikzlibrary{arrows.meta, positioning, shapes.geometric,calc, fit}
\usepackage{booktabs} % For better horizontal rules
\usepackage{siunitx}  % Optional: for aligning numbers/units
\usepackage{pifont}  % For the X mark (\multimap or \ding{55})

\definecolor{larakeyword}{RGB}{0,76,153}
\definecolor{laracomment}{RGB}{0,110,70}
\definecolor{larastring}{RGB}{153,51,0}
\definecolor{lararule}{RGB}{210,210,210}

\usepackage{xspace}
\newcommand{\RemoteManager}{\texttt{remotemanager}\xspace}
\newcommand{\LARA}{\textsc{LARA-HPC}\xspace}
\newcommand{\BigDFT}{\textsc{BigDFT}\xspace}

\newcommand{\MCP}{\textsc{MCP}\xspace}

\makeatletter
\AtBeginDocument{%
  \if@filesw
    \immediate\write\@auxout{\string\citation{apsrev42Control}}%
  \fi
}
\makeatother

\begin{document}

% \title{\LARA: Validation-Driven Agentic Workflows for Atomistic Modeling on HPC}
\title{\LARA: Validation-Driven Agentic Supercomputer Workflows for Atomistic Modeling}

\author{William Dawson}
\affiliation{RIKEN Center for Computational Science, Kobe, Hyogo 650-0047, Japan}

\author{Louis Beal}
\affiliation{INRIA, France}

\author{Yoann Cur\'e}
\affiliation{University Grenoble Alpes, MEM, L\_Sim, F-38000 Grenoble, France}

\author{Giuseppe Fisicaro}
\affiliation{Istituto per la Microelettronica e Microsistemi, Consiglio Nazionale delle Ricerche, Z.I. VIII Strada 5, I-95121 Catania, Italy}

\author{Dorian Rolland}
\affiliation{University Grenoble Alpes, MEM, L\_Sim, F-38000 Grenoble, France}

\author{Luigi Genovese}
\affiliation{University Grenoble Alpes, MEM, L\_Sim, F-38000 Grenoble, France}

\begin{abstract}

Large language models (LLMs) and agentic systems have recently demonstrated potential for automating scientific workflows, including atomistic simulations. However, their deployment in high-performance computing (HPC) environments remains limited by the lack of mechanisms ensuring correctness, reproducibility, and safe interaction with computational resources. Generated workflows suffer from inconsistencies, incorrect API usage, or invalid physical configurations --- leading to failed or unreliable simulations. In this work, we introduce \LARA, a validation-driven agentic framework to enable reliable workflow generation for atomistic modeling on HPC systems. Our approach is based on three key components: (i) a controlled execution layer that mediates all interactions with HPC resources; (ii) simulation-native validation through dry-run capabilities, enabling execution-level verification without incurring resource cost; and (iii) a multi-phase agentic pipeline combining retrieval-augmented generation and iterative refinement. We demonstrate the effectiveness of this approach performing an end-to-end atomistic simulation workflow on HPC by applying LARA-HPC to Density Functional Theory simulations. 
%BigDFT represents a massively parallel, wavelet-based first principle code - at density fuctional theory level - designed for high-performance computing and capable of scaling to thousands of heterogeneous nodes and targeting simulation at the exascale level.
%We demonstrate the effectiveness of this approach on representative simulation tasks using a prototype implementation that integrates a lightweight RAG mechanism based on BigDFT tutorial data and validation up to the dry-run stage. 
The results show that validation-driven generation significantly improves robustness and enables iterative correction of both syntactic and physical inconsistencies.
More broadly, this work advocates for a shift from generation-first to validation-first paradigms in Artificial Intelligence (AI) assisted scientific computing. We argue that the future task of the computational physics community is to develop domain specific agentic systems based on structured tooling to realize an HPC enabled co-piloted research ecosystem.

%We argue that structured HPC tooling and sandboxed environments are essential for integrating AI systems into production workflows, paving the way toward reliable co-piloted research ecosystems.

\end{abstract}

\maketitle

\section{Introduction}
\label{sec:introduction}

Recent advances in large language models (LLMs) have significantly expanded the capabilities of automated code generation, enabling the construction of increasingly complex workflows from natural language descriptions.
These models have demonstrated strong capabilities in reasoning over structured tasks, leading to the development of Artificial Intelligence (AI) copilots for software engineering~\cite{copilot2021,gottweis2025towards}. In scientific domains such as computational chemistry and materials science, these developments open new opportunities for automating simulation pipelines and accelerating research workflows~\cite{Zimmermann_2025,LLMHackathon2024,Alampara2026}.

\begin{figure}[t]
\centering
\includegraphics[width=\linewidth]{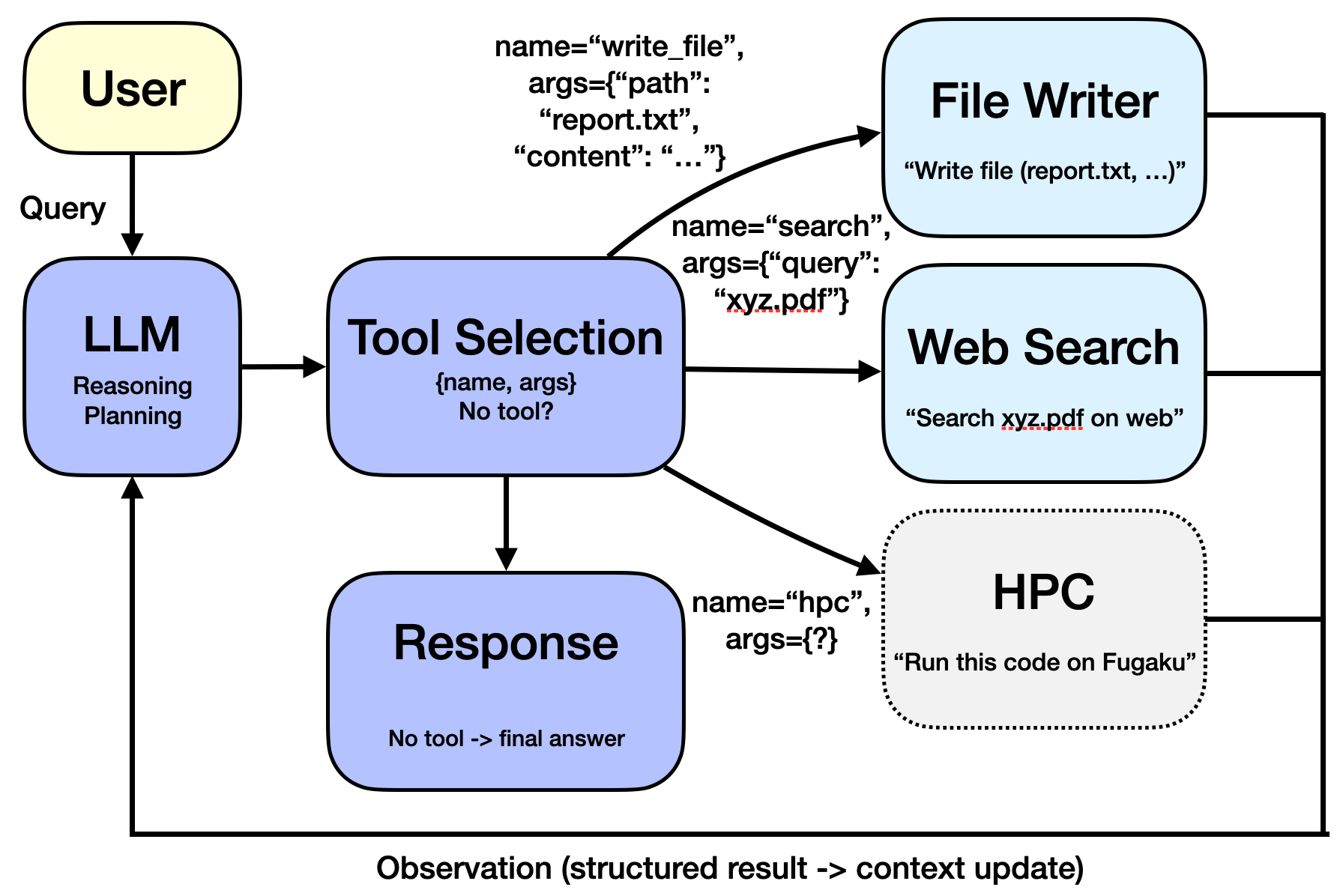}
\caption{ReAct (Reasoning and Acting) loop~\cite{yao2023react} at the foundation of agentic AI. The inner reasoning monologue of the LLM produces text that corresponds with specific tool calls. These tool calls perform actions, which result in structured observations being injected into the LLM's context for further reasoning. The missing piece addressed by this work is the integration of HPC.}
\label{fig:agentic_ai}
\end{figure}

In particular, LLM-based copilots and agentic systems (Fig.~\ref{fig:agentic_ai}) have demonstrated promising results in assisting scientific tasks, including experimental automation and atomistic simulations~\cite{mandal2025,bran2024,vriza2026,Lu2026-ks,gao2025democratizing,Qu2026,Zou2025}.
Several frameworks have been proposed to automate molecular dynamics workflows or computational chemistry pipelines~\cite{chemgraph2025,mdcrow2025,LLMHackathon2024}.
Agentic systems like DREAMS (DFT-based Research Engine for Agentic Materials Screening) aims at automating high-fidelity density functional theory (DFT) simulations through coordinated LLM agents, enabling autonomous, expert-level materials discovery and systematic error handling~\cite{wang2025dreams}.
Despite these advances, the deployment of such systems in high-performance computing (HPC) environments remains challenging. Scientific software stacks are often developed within specialized communities, evolve rapidly, and lack stable or fully documented interfaces.

Scientific simulations in HPC environments involve multiple stages, including input preparation, job submission, execution, and post-processing. While workflow management systems have been developed to automate these processes and ensure reproducibility~\cite{deelman2015,Huber2020}, integrating AI-driven systems into these environments introduces additional challenges, such as the need for controlled access to resources and compatibility with scheduling systems. %Traditional approaches based on direct scripting or manual interaction are not suitable for automated agents.
%As a consequence, 
LLM-generated workflows frequently suffer from inconsistencies, incorrect application programming interface (API) usage, or incomplete configurations. In HPC contexts, these issues are  critical as erroneous inputs lead to wasted computational resources, failed jobs, or scientifically invalid results. It is not sufficient to use a system with merely a high success rate (e.g., 95\%), as erroneous runs can have substantial consequences. HPC systems also often have high latency for job execution; hence, agentic systems cannot achieve productivity using a trial-and-error driven strategy, even if that is sufficient on a local workstation. In addition, the substantially higher number steps to handle is also a source of further difficulty in understanding the process, not only when something goes wrong, but also in terms of reproducibility of the results.

The difficulty is not limited to software engineering.
Consider the context of electronic structure calculations of materials based on the widely used Kohn-Sham Density Functional Theory~\cite{martin2020electronic} (DFT).
In atomistic modeling based on electronic structure calculations, 
a scientifically meaningful result depends on a coherent chain of modeling decisions:
how the system is built, the approximations used for electron interactions (choice of exchange--correlation functional), the treatment of atomic cores and spin, the numerical accuracy settings, and the analysis of the results must all be compatible with one another~\cite{brazdova2013}.
A workflow that is syntactically correct may still be scientifically wrong. This makes atomistic simulation 
particularly demanding for agentic AI --- correctness must be established at the level of the computational method, not only at the level of code generation.

Existing approaches primarily focus on improving generation quality through prompt engineering, retrieval-augmented generation (RAG), or iterative interaction loops~\cite{chemgraph2025,mdcrow2025}. 
Among other techniques, RAG has emerged for improving reliability by grounding generation in external knowledge sources~\cite{rag2020}. In scientific applications, RAG can leverage curated datasets such as training materials, documentation, and previous workflows to reduce hallucinations and improve consistency. While these strategies improve the relevance of the generated code, they do not provide guarantees of correctness prior to execution.  In particular, they cannot capture domain specific dynamic constraints related to simulation setup, resource allocation, or domain-specific validity.

This issue becomes especially visible when one moves beyond a fixed tool-calling scenario. A narrowly defined tool interface may be sufficient for a predetermined task, but realistic scientific exploration quickly requires methodological branching. 
For instance, during a DFT study, one may need to rethink the choice of model system, change the approximations used for the description of atoms' core electrons and nuclei (pseudopotential families), compare open- and closed-shell references, perform convergence studies, or introduce a new post-processing steps in response to intermediate results. Such adaptations are routine for expert practitioners, but are difficult to encode in advance through static tools.

In this work, we introduce \LARA, a validation-driven agentic framework designed to bridge this gap. The key idea is to shift from a \emph{generation-first} paradigm to a \emph{validation-first} approach (``Look Before You Leap''). 
This is achieved by combining: 1) a controlled HPC execution layer based on our Python \RemoteManager package to address operational and scheduling constraints; 2) simulation-native validation utilizing a ``dry-run'' primitive; and 3) a multi-phase agentic pipeline allowing for workflow generation and iterative refinement.

The dry-run primitive refers to running a computer program in a way that the code only validates the input and provides information about the process to be executed (an example of this feature is in the Linux utility \texttt{rsync}). Unlike static analysis, dry-run evaluation leverages the internal logic of the simulation code to detect errors in input configuration, resource requirements, or physical consistency, without the cost of a full calculation.
This enables iterative refinement in which workflows are progressively corrected until they satisfy all validation criteria and are ready for safe submission.
%Combining this approach with a controlled HPC execution layer and a multi-phase agentic pipeline, the dry-run primitive enables an iterative refinement loop in which workflows are progressively corrected until they are ready to be safely submitted on an HPC resources.

We showcase this approach by executing full atomistic simulation workflows on HPC, applying LARA-HPC to orchestrate DFT based simulations with the \BigDFT program~\cite{ratcliff2020}. BigDFT is an ab-initio electronic structure code. It is built for extreme-scale HPC systems, using linear-scaling algorithms and efficient parallelization to simulate very large systems, making it relevant for exascale computing architectures.
Two representative atomistic simulation tasks will be presented: 1) the calculation of the atomization enthalpy of a molecular system and 2) surface adsorption energies of a large system. Our experiments demonstrate how an agentic system incorporating dry-run validation and structured feedback can identify incorrect API usage, missing inputs, and inconsistent physical settings before production execution. We argue that such validation-driven paradigms are essential for ensuring correctness, reproducibility, and efficient use of computational resources in future copiloted research. 

\section{HPC Orchestration with \RemoteManager and Model Context Protocol}
\label{sec:remotemanager-mcp}

Before introducing \LARA itself, it is important to describe the execution substrate on which the framework relies. In our view, the first challenge in connecting agentic systems to atomistic simulation campaigns is not code generation, but the controlled exposure of HPC resources. An AI system should not directly manipulate login shells, schedulers, filesystem paths, or ad hoc environment configuration. Instead, these interactions should be mediated through a reproducible and auditable orchestration layer.
For this purpose, we build on \RemoteManager~\cite{dawson2024}, a lightweight Python middleware designed to promote local Python functions into remotely executable tasks on HPC resources. The key design principle of \RemoteManager is to preserve the standard user workflow of scientific computing --- batch scripts, file staging, remote execution, and result retrieval --- while expressing it through a programmatic interface. In this way, the library does not replace nor circumvent the logic of HPC usage, but formalizes it.

The core abstraction of \RemoteManager is that an ordinary Python function can be \emph{promoted} into a remote procedure call enveloped in a user defined execution environment (executables, libraries, data, etc).
% The core abstraction of \RemoteManager is that an ordinary Python function can be \emph{promoted} into a remote callable.
The user first specifies a jobscript template adapted to the target machine. This template contains the scheduler directives and execution environment, but also exposes a controlled set of tunable parameters such as the number of nodes, MPI ranks per node, and OpenMP threads. Listing~\ref{lst:template} illustrates this first step.

\begin{lstlisting}[caption={A jobscript template of supercomputer Fugaku (a petascale supercomputer at the RIKEN Center for Computational Science in Kobe, Japan) for use with \RemoteManager. Optional parameters such as the number of nodes, MPI ranks per node, and OpenMP threads are explicitly exposed. The jobscript defines the execution environment by loading modules, running containers, setting environment variables, etc (in this case, through a conda virtual environment).}, label={lst:template}]
template = """#!/bin/bash -l
#PJM --mpi `max-proc-per-node=#MPI#'
#PJM -L `elapse=2:00:00'
#PJM -L `node=#NODES#'
#PJM -L `rscgrp=small'
#PJM -g ra000000

export OMP_NUM_THREADS=#OMP#

# Example of environment initialization on the target resource
eval "$($DATADIR/bin/micromamba shell hook --shell bash)"
micromamba activate remote"""
\end{lstlisting}

This template-based design is important for two reasons. First, it reflects how production HPC jobs are actually launched: the scheduler remains the entry point, rather than an opaque service layer. Second, it makes resource selection explicit and inspectable. The agent or higher-level workflow may choose values for exposed parameters, but it cannot arbitrarily redefine the structure of the batch script.

Once the template is defined, the user instantiates a \RemoteManager computer object, completes the machine-specific configuration, and promotes 
the Python function (Listing~\ref{lst:remote}).
%a Python function into a remote callable.
%This process is shown in Listing~\ref{lst:remote}. 
The promoted function can then be invoked from the local Python session as if it were a normal function, while \RemoteManager transparently handles file transfer, job submission, remote execution, and result collection.

\begin{lstlisting}[language=Python, caption={Configuring a remote machine and promoting a BigDFT single-point calculation of H$_2$O for remote execution on supercomputer Fugaku.}, label={lst:remote}]
Fugaku = Computer(template=template,
                  host="fugaku.r-ccs.riken.jp",
                  submitter="pjsub")
Fugaku.nodes = 1
Fugaku.mpi = 1
Fugaku.omp = 48

@SanzuFunction(url=Fugaku)
def run_water_energy(hgrid=0.4, dry_run=False):
    from BigDFT.Systems import System
    from BigDFT.Fragments import Fragment
    from BigDFT.Atoms import Atom
    from BigDFT.Inputfiles import Inputfile
    from BigDFT.Calculators import SystemCalculator

    sys = System()
    sys["WAT:0"] = Fragment([
        Atom({"O": [0.0, 0.0, 0.0], "units": "bohr"}),
        Atom({"H": [1.43, 1.11, 0.0], "units": "bohr"}),
        Atom({"H": [1.43, -1.11, 0.0], "units": "bohr"}),
    ])

    inp = Inputfile()
    inp.set_xc("PBE")
    inp.set_hgrid(hgrid)

    code = SystemCalculator(dry_run=dry_run)
    log = code.run(sys=sys, input=inp, name="water")
    return getattr(log, "energy", None)

run_water_energy(hgrid=0.35, dry_run=True)
\end{lstlisting}
Conceptually, the promoted function defines the \emph{scientific payload}, while \RemoteManager supplies the \emph{execution protocol}. This separation is particularly useful in atomistic simulations, where scientific code such as \BigDFT may evolve rapidly but the operational steps of HPC execution remain comparatively stable.

\subsection{Operational role of \RemoteManager}

\label{subsec:operational}

At execution time, \RemoteManager carries out the sequence of actions that a human user would otherwise perform manually. These typically include:
\begin{enumerate}
    \item materializing a batch script from the selected template,
    \item staging input files and Python sources to the remote resource,
    \item submitting the job to the scheduler,
    \item monitoring completion,
    \item retrieving output files, return values, and execution logs.
\end{enumerate}
A major strength of this approach is that it does not require any daemon or privileged service on the remote machine. The remote side only needs the standard tools already available to HPC users, such as a shell environment, scheduler commands, and secure file transfer. This makes the approach portable across production systems and compatible with the security constraints of real HPC centers. The design also means that \RemoteManager is not tied to any particularly job scheduling software. 
Equally important is the fact that \RemoteManager captures not only the nominal return value of the remote function, but also standard output, standard error, and execution artifacts. This is essential in scientific settings, where failed runs are often as informative as successful ones, and where debugging requires visibility into the exact job context.

\subsection{Towards an HPC tool call}
While \RemoteManager provides the execution substrate, it does not by itself define how an LLM or AI agent should access these capabilities. For that interface layer, we use the Model Context Protocol (\MCP), which standardizes the exposure of structured tool calls to large language models.
In a typical \MCP server based on Python, a function is annotated and published as a tool. The function signature and docstring become part of the semantic interface presented to the model. When this mechanism is combined with \RemoteManager, an \MCP tool call can be mapped onto a remote HPC submission. Listing~\ref{lst:combo} illustrates this composition.

\begin{lstlisting}[language=Python, caption={Exposing the same H$_2$O single-point routine as an \MCP tool while keeping the remote resource configuration explicit. Here we use the FastMCP library (gofastmcp.com).}, label={lst:combo}]
mcp = FastMCP("Example Server")

@mcp.tool()
def water_single_point(hgrid, dry_run, omp, mpi, nodes):
    """
    Run a BigDFT single-point calculation for H2O.
    Args:
      hgrid: real-space grid spacing.
      dry_run: use BigDFT dry-run validation mode.
      omp: number of OpenMP threads to use.
      mpi: number of MPI ranks per node.
      nodes: number of nodes to use.
    """
    Fugaku.nodes = nodes ; Fugaku.mpi = mpi ; Fugaku.omp = omp
    return run_water_energy(hgrid=hgrid, dry_run=dry_run)
\end{lstlisting}

This functionality composition is the key bridge between agentic reasoning and production HPC campaigns. The LLM does not receive shell access, scheduler access, or arbitrary filesystem access. Instead, it sees a constrained set of tools with well-defined parameters. Each invocation is then converted into an auditable batch submission by \RemoteManager.

\subsection{Strengths and limitations of the \MCP approach}
The \MCP with \RemoteManager combination already solves a substantial part of the AI--HPC integration problem. It provides a secure and realistic path for exposing supercomputing resources to an external reasoning system without compromising the operational model of the target platform. However, this approach also reveals a structural limitation. 
Each exposed \MCP tool corresponds to a predefined function and therefore to a predefined workflow pattern. This is suitable for repetitive or narrowly scoped tasks, but it becomes restrictive when the scientific workflow must adapt dynamically. One solution would be to provide an agentic framework with a broad range of built in tools~\cite{Zou2025}, however even this may fail to cover the dynamic use cases pursued by expert simulation scientists.

% to me this is redundant
%In particular, it offers three major advantages.

%First, it enforces clear security boundaries: only explicitly defined tools may be invoked. Second, it preserves portability, since the remote execution details remain encapsulated in the \RemoteManager configuration rather than scattered across agent prompts or tool definitions. Third, it aligns naturally with production workflows, because jobs are still launched through scheduler-backed templates and standard remote operations.

%However, this approach also reveals a structural limitation. 
%Each exposed \MCP tool corresponds to a predefined function and therefore to a predefined workflow pattern. This is suitable for repetitive or narrowly scoped tasks, but it becomes restrictive when the scientific workflow must adapt dynamically to intermediate results, new hypotheses, or alternative software configurations. In atomistic modeling, where one frequently needs to change numerical parameters, add preparation steps, or introduce new post-processing logic, static tool definitions quickly become too rigid.

This limitation may be stated more concretely through the example of molecular atomization energies. Computing an atomization energy is conceptually simple, but the workflow requires coordinated calculations for the molecule and its isolated constituent atoms, with a consistent choice of exchange--correlation functional, spin configuration, boundary conditions, numerical parameters, and post-processing conventions. Small methodological changes, such as switching pseudopotential families or tightening convergence parameters, can alter the resulting quantity appreciably.

Within an \MCP-based framework, one could expose a tool for building a molecular system, another for running a \BigDFT calculation, and a final one for aggregating energies. Such a setup is sufficient for a single pre-planned workflow. But as soon as the scientific question evolves to include alternative pseudopotentials, open-shell atomic references, all-electron variants, or convergence scans over grid spacing and localization radii, the number of required tool variants grows rapidly. The agent can only choose among the interfaces that were explicitly designed in advance; it cannot introduce a new methodological branch on its own.

For this reason, we regard the \MCP plus \RemoteManager layer 
%not as the final solution, 
not as a complete solution,
but as the indispensable execution backbone on top of which a more flexible system can be built. In the next section, we introduce \LARA, which retains the controlled HPC interaction offered by \RemoteManager while replacing static workflow definitions by validated, retrieval-guided code generation.

\section{LARA-HPC Architecture}
\label{sec:lara-architecture}
The central limitation of the HPC tools approach, as discussed in the previous section, is its reliance on predefined actions. While this ensures safety and reproducibility, it restricts the ability of AI agents to construct workflows dynamically. 
In practice, simulation based research often requires the composing of new steps and logic to encompass and expanding set of hypotheses and scientific questions.
%In practice, scientific simulations often require adapting parameters, composing multiple steps, or introducing new logic depending on intermediate results. 
This motivates the introduction of a higher-level reasoning layer capable of generating and refining workflows while preserving the controlled interaction with HPC resources.

Our architectural choice is guided by a principle of minimal sufficient autonomy~\cite{blount2025agents}: the agent should be as flexible as needed for scientific reasoning, but no more autonomous than is operationally safe in an HPC setting. A fully deterministic pipeline is too rigid for open-ended scientific requests, while a fully autonomous agent with direct execution authority would be difficult to justify in a high-consequence environment where an incorrect job may waste substantial computational resources~\cite{wang2024agentsurvey,yang2024risks}. \LARA therefore adopts a staged design in which creative reasoning, deterministic validation, and controlled execution are explicitly separated. This positioning is also consistent with recent practical guidance on agent construction~\cite{openai2024agents,blount2025agents}, improving, in passing, also the traceability of the approach.

\subsection{Overview of the LARA Workflow}

\begin{figure}[t]
\centering
\includegraphics[width=\linewidth]{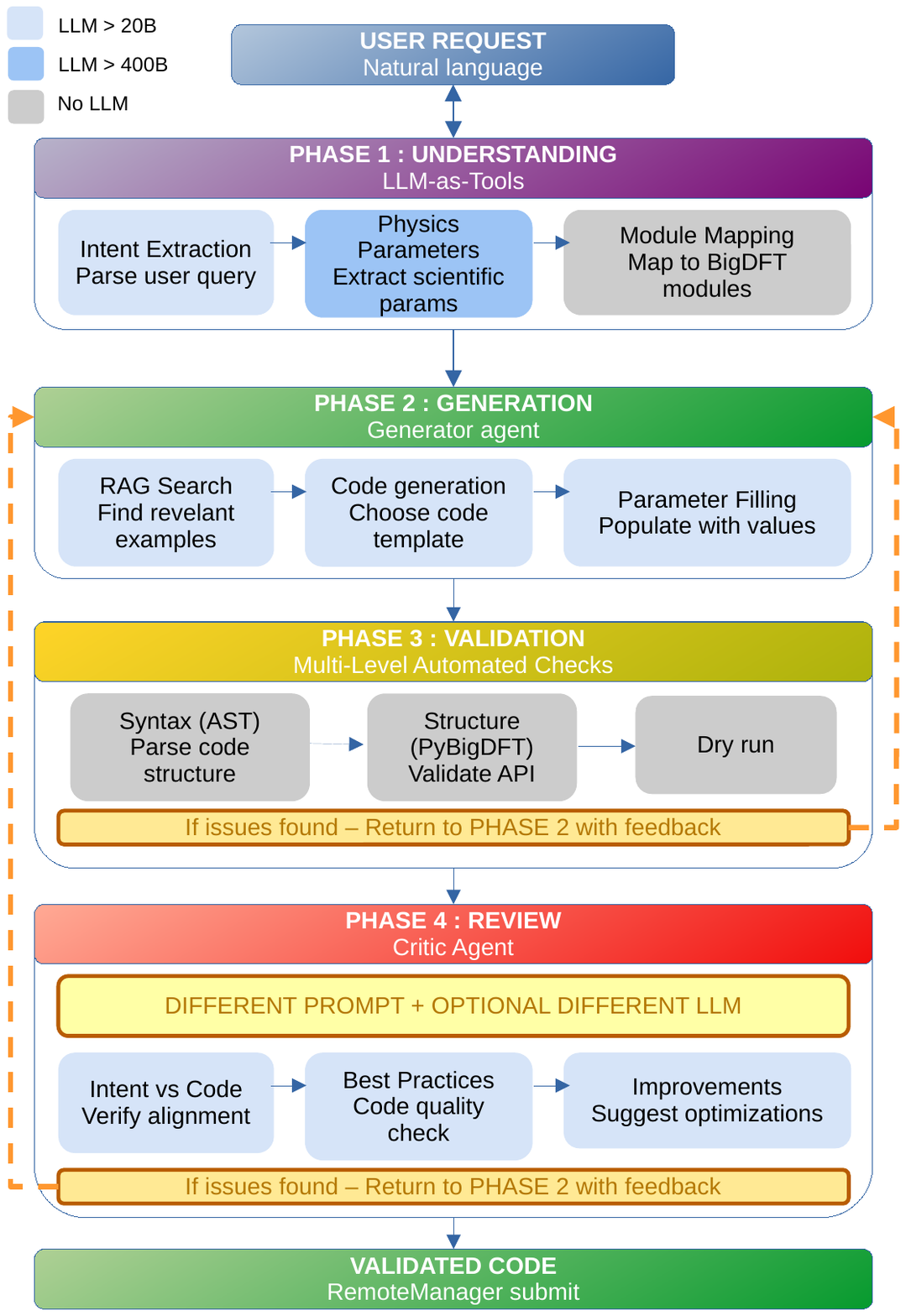}
\caption{LARA-HPC architecture. A user request is transformed through successive phases of understanding, generation, validation, and review, before producing validated code executed via \RemoteManager.}
\label{fig:lara_architecture}
\end{figure}

\LARA is structured as a multi-phase pipeline that transforms a natural language request into a validated and executable workflow. The architecture is illustrated in Fig.~\ref{fig:lara_architecture}.
The workflow is decomposed into four main phases:
\begin{enumerate}
    \item \textbf{Understanding}: extraction of scientific intent and mapping to domain-specific modules.
    \item \textbf{Generation}: construction of candidate workflows using templates and retrieval.
    \item \textbf{Validation}: multi-level verification including syntax, API, physical constraints, and dry-run execution.
    \item \textbf{Review}: higher-level critique and optimization of the generated workflow.
\end{enumerate}
This structured decomposition allows separating concerns between semantic interpretation, code generation, correctness verification, and optimization.

The decomposition also reflects distinct failure modes. Misunderstanding the user request leads to an inappropriate workflow family; hallucinations during generation lead to incorrect code or invalid API usage; subtle scientific inconsistencies survive syntax checks but are exposed by domain validation; and workflows that are formally correct may still be suboptimal or misaligned with the original intent. The four-phase structure is intended to address each of these failure modes with a dedicated mechanism rather than a single monolithic agent prompt. A similar phasing strategy has been adopted by the agentic system ``The AI Scientist'' --- a pipeline that achieves the vision of full end-to-end automation of the scientific process~\cite{Lu2026-ks}. The AI Scientist sequentially completes four main phases covering idea generation, tree-based experimentation, manuscript writing and reviewing.

A further design principle of LARA-HPC is the separation of cognitive labor across models of different sizes (highlighted by different colors in Fig.~\ref{fig:lara_architecture}). The most demanding reasoning step --- translating a scientific problem into physically meaningful parameters, understanding domain equations, and selecting the right computational strategy --- is handled by a single large model. This frontier model brings the broad scientific knowledge and deep reasoning capacity required to interpret quantum chemistry or electronic-structure problems correctly. Leveraging different LLM models at various stages and for different agent tasks is a common practice in agentic systems to best utilize specific skills (reasoning, coding, vision-language tasks) and to optimize costs~\cite{Lu2026-ks}. A model with strong tool calling capabilities can also be utilized to dynamically orchestrate the four stage workflow. Once this high-level physical reasoning is complete and the simulation parameters are established, subsequent well-scoped subtasks steps are delegated to smaller, faster, and cheaper models. 

%These lightweight agents handle well-scoped sub-tasks: formatting input files, invoking HPC job submission tools, validating syntax, or post-processing outputs. Because each of these tasks is narrow and clearly defined by the structured artifacts produced upstream, a small model is entirely sufficient.

%This asymmetric architecture yields significant practical benefits. It keeps inference costs low and latency manageable while preserving scientific correctness where it matters most. It also makes the system easier to deploy in resource-constrained environments — for instance, running the small agents locally while querying the large model via API only once per simulation. The result is a workflow that is both accurate and economically scalable.

\subsection{Phase 1: Scientific Understanding}
The first phase transforms the user request, expressed in natural language, into structured scientific information. This includes extraction of intent (e.g., energy calculation, structural relaxation), identification of relevant physical parameters, and mapping to available modules in the target simulation framework (e.g., \BigDFT). This step is critical because scientific workflows cannot be constructed solely from syntactic patterns: they require a correct interpretation of domain-specific concepts. The output of this phase is a structured representation of the problem, which serves as input for workflow generation.

In practical deployments, this phase also provides a natural location for clarification when the user request is underspecified. For example, an atomization-energy calculation may require the system to determine whether geometry optimization is intended, which pseudopotential family should be used, or whether open-shell reference atoms must be considered explicitly. Resolving such ambiguities before generation prevents avoidable downstream errors, and is consistent with recent guidance on human-in-the-loop agent design~\cite{wang2024agentsurvey,blount2025agents}.

\subsection{Phase 2: Workflow Generation}
The generation phase constructs candidate workflows by combining retrieval and templating mechanisms. It consists of three main steps:

\begin{itemize}
    \item \textbf{Retrieval (RAG)}: relevant examples and documentation are retrieved from a curated knowledge base.
    \item \textbf{Code generation}: a candidate function is generated according to a specific code template.
    \item \textbf{Parameter filling}: the source code is incorporated into a structured message incorporating output from the previous stages.
    % \item \textbf{Template selection}: a suitable code structure is selected based on the task.
    % \item \textbf{Parameter filling}: the template is instantiated with values derived from the user request.
\end{itemize}

This approach ensures that generated workflows remain anchored to known usage patterns, reducing the likelihood of hallucinated or inconsistent code.
A common approach for grounding LLM responses in domain-specific knowledge is retrieval-augmented generation, where relevant documentation chunks are fetched at inference time and injected into the model's context. This strategy works well when the knowledge base is large and dynamic, as it avoids overwhelming the context window with irrelevant information.
Another viable solution is represented by always-on memory (see Ref.~\cite{packer2024memgptllmsoperatingsystems,xu2025amemagenticmemoryllm} for discussions of agent memory). In this approach, domain tutorials are manually ingested into the model's persistent memory, tested against representative tasks, and refined until the model consistently produces correct results. This validated memory is then deployed as a static knowledge context throughout the workflow. 
%In a scientific setting where the relevant knowledge corpus is bounded and stable — such as the input conventions and computational parameters of a specific electronic-structure code — AOM is a natural fit: the memory is compact, verified, and always available without any retrieval step, making it both simpler to deploy and easier to audit.

However, at this stage, correctness is not guaranteed. The generated workflow is only a candidate that must be validated before execution.
This phase is intentionally more expressive than static tool-calling. It allows the system to construct new workflow logic when required by the scientific question, for example by adding atomic reference calculations, convergence loops, or parameter comparisons that were not hard-coded as independent tools.

\subsection{Phase 3: Multi-Level Validation}
The validation phase is the core innovation of \LARA. It introduces a layered verification process that progressively ensures the correctness of the generated workflow.
The role of validation is therefore not to replace generation, but to make this added expressivity safe enough for scientific use. In this sense, the architecture is closer to an agentic RAG workflow than to a static retrieval pipeline~\cite{singh2025agenticrag}.
The validation pipeline includes:
\begin{itemize}
    \item \textbf{Syntactic validation}: parsing of the abstract syntax tree (AST).
    \item \textbf{Structural validation}: verification of API usage (e.g., through BigDFT's Python interfaces).
    % \item \textbf{Physical validation}: enforcement of domain-specific constraints.
    \item \textbf{Dry-run execution}: evaluation of the workflow and its resource requirements using simulation code validation modes.
\end{itemize}
The dry-run step is particularly important, as it enables the detection of runtime and semantic errors without incurring the cost of full HPC execution.
If issues are detected, structured feedback is generated and the workflow is returned to the generation phase for correction. This creates an iterative refinement loop that continues until the workflow satisfies all validation criteria.

\subsection{Phase 4: Critical Review}
In addition to automated validation, \LARA includes a review phase driven by a dedicated agent. This phase operates with a potentially different prompt or even a different model, and focuses on higher-level aspects, such as alignment between user intent and generated code, adherence to best practices or identification of possible optimizations.
This step is not strictly required for correctness, but improves the overall quality and robustness of the workflow. It introduces a second level of reasoning that complements the validation phase.
Although the present version does not yet implement this phase end-to-end, we keep it in the architecture because scientific acceptability is broader than executability. A workflow can pass dry-run validation and still be methodologically weak, unnecessarily expensive, or poorly aligned with the actual question being asked. The review phase is intended to capture precisely this distinction, and follows the broader generator-critic pattern explored in recent self-refinement work~\cite{madaan2024selfrefine}.

\subsection{Key Design Principles}
Once a workflow has successfully passed validation and review, it is considered safe for execution. At this stage, the code is submitted to the HPC environment through \RemoteManager.
% not an important detail, and inconsistent with the sanzu api we showed earlier
%typically via a \texttt{.run()} call (the main method of a \RemoteManager orchestrated function)
This final step closes the loop between agentic reasoning and production execution. Importantly, the agent never interacts directly with the HPC system: all operations are mediated through the \RemoteManager layer described in Section~\ref{sec:remotemanager-mcp}. Everything happens on the local user's workstation.

The architecture of \LARA is guided by three key principles:
\begin{itemize}
    \item \textbf{Separation of concerns}: reasoning, validation, and execution are handled by distinct components.
    \item \textbf{Validation-first design}: workflows must be verified before execution.
    \item \textbf{Controlled interaction}: all HPC operations are mediated through well-defined interfaces.
\end{itemize}
These principles ensure that flexibility in workflow generation does not compromise the reliability and reproducibility required for HPC-based scientific computing.

\section{Dry-Run Driven Validation}
\label{sec:dryrun}

A central component of the \LARA framework is the integration of a \emph{dry-run} execution mode, which enables the validation of simulation workflows prior to their actual execution on HPC resources. Unlike traditional static validation approaches, the dry-run mechanism in \BigDFT is implemented directly within the execution pipeline of the simulation code, allowing for domain-aware verification of inputs.
In \BigDFT, the dry-run mode can be triggered in two complementary ways:
\begin{itemize}
    \item through the Python API, by calling the \texttt{run()} method of the calculator with the keyword argument \texttt{dry\_run=True},
    \item through the environment, by setting the variable \texttt{BIGDFT\_MPIDRYRUN}, which redirects execution to a dedicated dry-run executable. The value of the variable corresponds to the number of MPI processes the simulation will use.
\end{itemize}

In both cases, the dry-run mode modifies the execution pathway of the calculation. 
% we don't show anything so this is not clear
%As shown in the implementation of the \texttt{SystemCalculator} class, the command associated with the execution is dynamically constructed depending on the value of the \texttt{dry\_run} flag.
When dry-run is activated, the standard \BigDFT executable is replaced by a specialized tool
%\begin{lstlisting}[language=bash]
%bigdft-tool -a memory-estimation -l
%\end{lstlisting}
%This command performs a simulation of the execution, including input parsing and internal setup, but does not allocate the computational resources required for a full self-consistent field (SCF) calculation.
that performs a simulation of the execution, including input parsing and internal setup, but does not allocate the computational resources required for a full self-consistent field (SCF) calculation.

The dry-run mechanism is not an external checker, but is embedded within the same execution pipeline used for actual simulations. It thereby provides a rich set of validation capabilities that go beyond simple syntactic checks. These include:
\begin{itemize}
    \item \textbf{Input validation}: parsing and verification of input files.
    \item \textbf{Structural consistency}: validation of parameter dictionaries and required fields.
    \item \textbf{Physical setup}: construction of internal data structures such as grids, basis functions, and pseudopotentials.
    \item \textbf{Resource estimation}: evaluation of memory requirements based on system size and parallel configuration.
    %\item \textbf{Software}: validation of the loaded software enviroment, code installation and proper executable call. % GIUSEPPE: It is worth to a check of code directly in HPC
\end{itemize}
In particular, the memory estimation step provides a quantitative assessment of the feasibility of the simulation under given HPC conditions. The resulting logs contain a \emph{Memory Consumption Report}, which is explicitly checked in the calculator implementation to confirm that the dry-run completed successfully.
This level of validation enables the detection of errors that would otherwise only appear during execution, such as inconsistent parameter combinations or infeasible resource requirements. It further provides feedbacks to the agentic system concerning the optimal allocation of HPC resources.

The dry-run execution produces standard output and log files that are processed in the same way as full runs. The calculator inspects debug files and log timestamps to determine whether the execution was successful. In case of failure, detailed error information is retrieved from dedicated debug files and reported back to the user or calling system.
This mechanism provides structured feedback that can be directly used by the \LARA validation loop to refine the workflow. Unlike generic error messages, these diagnostics are grounded in the internal logic of the simulation code, making them particularly valuable for automated correction.

A key advantage of the dry-run mode is its low computational cost. Since no full SCF cycle is performed, the execution time is negligible compared to a production run, even for large systems. This makes it possible to invoke dry-run repeatedly within an iterative loop without impacting HPC resource usage.
This property is essential for \LARA, where validation is performed multiple times during the generation-refinement cycle. The dry-run thus acts as a fast and reliable oracle for workflow correctness.

Within the \LARA architecture (Section~\ref{sec:lara-architecture}), the dry-run constitutes the final stage of the validation phase. It complements earlier checks such as syntax parsing and API validation by providing an execution-level verification grounded in the simulation code itself.
This integration enables a shift from \emph{static validation} to \emph{simulation-native validation}. Instead of attempting to predict correctness from code structure alone, \LARA relies on the simulation engine to evaluate the validity of the workflow.
%This design is fundamentally different from standard AI-assisted coding approaches. It ensures that generated workflows are not only syntactically correct, but also physically meaningful and executable in a real HPC environment.
As a result, only workflows that pass the dry-run validation are submitted to execution through \RemoteManager, guaranteeing a high level of reliability in production campaigns.

\section{Use Case: Validation-Driven Workflow Generation}
\label{sec:usecase}

To illustrate the capabilities of the \LARA framework, we present concrete use cases (atomization energy and surface binding) based on an implementation of the architecture described in Section~\ref{sec:lara-architecture}. This implementation focuses on the generation and validation of atomistic simulation workflows using \BigDFT, and is designed as a minimal yet representative instance of the overall architecture.

The implementation considered in this work realizes the \LARA pipeline up to \textbf{Phase 3 (Validation)} of Fig.~\ref{fig:lara_architecture}.
% I change this from curated corpus because I think the reader should get that our documentation is not really that great
The RAG component is implemented using materials from a \BigDFT training event,
%The RAG component is implemented using a curated corpus derived from the \emph{BigDFT-school tutorials},
which provide representative examples of input configurations and workflows. 
This choice ensures that generated code remains consistent with established usage patterns while keeping the implementation lightweight.
%The present paper therefore reports a prototype rather than a complete autonomous scientist.
We use Google's GEMINI 2.5 Flash~\cite{comanici2025gemini25pushingfrontier} as the smaller, stage-specific LLM; \textit{gemini-embedding-001} as the embedding model; and Anthropic's Sonnet 4.6 as the frontier model. This section demonstrates that scientific workflow generation can be coupled to simulation-native validation in a way that catches meaningful errors before production execution. The goal of the use case is not to claim exhaustive automation, but to show that the validation loop materially changes what kinds of workflows can be trusted.

\subsection{Illustrative Error Detection and Correction}
To provide a concrete understanding of the validation-driven workflow, we present representative examples of errors generated during the atomization workflow construction and their subsequent correction through the \LARA pipeline.

\paragraph{API Misuse}
A common class of errors arises from incorrect usage of the simulation API.
For instance, the generated workflow may import a nonexisting/mispelled Python module.
During the validation phase, this error is detected either at the structural validation level or during dry-run execution, producing an error message indicating that the module does not exist.
Another concrete example %from the prototype
concerns dynamic self-correction after an \texttt{AttributeError}. Starting from a natural-language request to construct a water molecule in a periodic box and return its center of mass, the agent first produced an invalid call to \texttt{System.get\_center\_of\_mass()}. The dry-run stage intercepted the error, after which the generation phase consulted the indexed documentation and replaced the invalid method by the valid \texttt{System.centroid} attribute. The corrected workflow then executed successfully and returned a coordinate vector. This example shows the intended control loop: failure is not hidden, but converted into structured feedback that triggers a focused repair.

\paragraph{Missing Input Parameters}
Another frequent issue concerns incomplete or inconsistent input definitions, such as missing keys in the input data.
The dry-run execution detects missing required fields (e.g., atomic positions or pseudopotentials), leading to a failure at the input parsing stage.

\paragraph{Physically Inconsistent Parameters}
More subtle errors involve inconsistent configurations. A representative example is a spin-polarized calculation on a single hydrogen atom with an impossible total magnetic moment:
%\begin{lstlisting}[language=Python, caption={Physically inconsistent configuration}]
\begin{lstlisting}[language=Python]
from BigDFT.Database.Molecules import get_molecule
from BigDFT.Inputfiles import Inputfile

sys = get_molecule("H")
inp = Inputfile()
inp.set_xc("PBE")
inp.set_hgrid(0.4)
inp.spin_polarize(2)
\end{lstlisting}
This configuration is syntactically valid Python and uses the correct API, but it violates a physical constraint because a one-electron system cannot sustain a total spin polarization of two. In the executable validation path, BigDFT reports the issue explicitly:
%\begin{lstlisting}[caption={Diagnostic reported by BigDFT for the invalid hydrogen spin configuration}]
\begin{lstlisting}
ERROR: some problem occured during the execution of the command
The error occured is BIGDFT_INPUT_VARIABLES_ERROR
Additional Info: Spin-Polarized calculation (nspin= 2).
The mpol polarization should be less than the number of electrons
(mpol=2 and nel= 1)
\end{lstlisting}
The corrected workflow simply restores a physically admissible polarization.
%for example:
%\begin{lstlisting}[language=Python, caption={Corrected physically consistent configuration}]
%\begin{lstlisting}[language=Python]
%inp.spin_polarize(1)
%\end{lstlisting}

These examples highlight the multi-level nature of the \LARA validation process. Errors are not only syntactic but also structural and physical. The dry-run mechanism plays a central role in detecting issues that cannot be inferred from code inspection alone.
By iteratively correcting such errors, the system converges toward workflows that are both syntactically valid and scientifically meaningful. This illustrates the transition from a generative paradigm to a validation-driven construction process.
The spin polarization failure raises at the same time also a general crucial point: \LARA relies on the robustness of a code with respect to exceptions handlings and error descriptions. \LARA was able to set the proper spin for hydrogen since \BigDFT makes users aware of the proper correlation of spin state and available electrons.
This example point to the need of having a \emph{specific} dry run tool, designed by the developers of the code into which the HPC campaign should run, that will be consumed by the agent to meet required validation criteria.
The ability of the software developer will be then in being capable to design such a tool in a suitable way to inform HPC campaigns and relevant worflows actions.

\subsection{Why Atomization Energy Matters as a Use Case}

Beyond these local error-correction examples, the broader scientific motivation for \LARA is the handling of workflows whose difficulty lies in methodological branching rather than in syntax alone. Atomization energies provide a compact example. To compute them correctly, the agent must coordinate molecular and atomic calculations, preserve the same exchange--correlation and numerical settings across all runs, choose appropriate reference spin states for isolated atoms, and often explore variants such as pseudopotential family or convergence parameters.

In a static tool-calling framework, each of these scientific variants require an explicit new tool or a more complex interface. In a generation-based framework, the agent can instead synthesize the required workflow structure on demand. The price of this flexibility is that the resulting code must be validated much more carefully. For this reason, atomization energy is not merely a chemistry example --- it is a representative stress test for the central claim of this paper, namely that scientific usefulness requires more expressivity than fixed tools, but HPC safety requires more discipline than unconstrained code generation.

Using \LARA to generate a code that converges the atomization energies revealed another limitation of a predefined tool approach. \BigDFT represents the electronic structure using wavelets, which support a multiresolution approach. Around each atom, a coarse-grained region with lower resolution reduces computational cost, and its size can be adjusted via the \texttt{set\_rmult()} function.
The key point is not the specific parameter, but the fact that the system had to modify workflow logic rather than merely fill slots in a pre-existing interface (Listing~\ref{lst:rmult}).

\begin{lstlisting}[language=Python, caption={Excerpt of a validated code modification generated by \LARA when extending a BigDFT workflow to accept \texttt{crmult} through \texttt{set\_rmult()}.}, label=lst:rmult]
def input_template(xc='PBE', hgrid=0.35, **kwargs):
    from BigDFT.Inputfiles import Inputfile
    inp0 = Inputfile()
    inp0.set_xc(xc)
    inp0.set_hgrid(hgrid)
    inp_spec = {'XC': xc, 'h': hgrid}

    if 'crmult' in kwargs:
        crmult_value = kwargs.pop('crmult')
        inp0.set_rmult(crmult_value)
        inp_spec['rmult'] = crmult_value

    return inp0, inp_spec
\end{lstlisting}

One possible solution is to create a fully general interface exposing all possible input options. Another would be to allow the agent to first generate an input file that is sent to the MCP server. Both of these solutions place a heavy burden on prompt engineering to ensure the agent properly manages an enormous array of possible input options and all of their dynamic constraints. Thus, the \LARA approach is to enable the generation of validated, programmatic workflows in the context of the most relevant documentation and physics considerations. Once the workflow passes all validation steps, it can be safely executed on HPC resources using the \RemoteManager interface described in Section~\ref{sec:remotemanager-mcp}.
At this stage, the workflow is no longer a speculative generated code, but a verified simulation procedure that has been tested under realistic conditions.

\subsection{Validation Driven HPC}

To connect AI agents to HPC resources for predefined workflows, it is sufficient to use \RemoteManager as glue code through the scheme in Listings~\ref{lst:remote} and~\ref{lst:combo}. However, the purpose for \LARA is to enable autonomous AI assistants to run novel, computationally demanding calculations having validated the workflow locally. As a demonstration of this point, the orchestrating agent is prompted with the following query:
\begin{lstlisting}
Inside the data directory there is a XYZ file. I want to
calculate the binding energy of the water molecule to the
surface in that geometry using the PBE functional.
\end{lstlisting}
% As in the previous subsection, \LARA is used to generalize the previous workflow on atomization energies to one for computing surface adorption energy:
% \begin{equation}
% E_{\mathrm{ads}} = E_{\mathrm{combined}} - E_{\mathrm{surface}} - E_{\mathrm{adsorbate}}.
% \end{equation}
In this scenario, the LLM is provided with an anatase/water structure composed of 291 atoms~\cite{Fisicaro-JPCC-2020}. 

\begin{table*}[t]
    \caption{Memory analysis produced by the AI agent for calculating the slab system. The subcaptions, headers, and entries are all verbatim taken from the agent's output.}
    \label{tab:memory}

    \centering
    (a) Memory estimates per component (1 process dry-run)\par\vspace{0.5em}
    \begin{tabular*}{\textwidth}{@{\extracolsep{\fill}}lll}
        \toprule
        \textbf{Calculation} & \textbf{Wavefunction Memory} & \textbf{Estimated Peak} \\
        \midrule
        slab + water (291 atoms)  & $\sim$1784 MB & $\sim$197 GB \\
        bare slab (288 atoms)     & $\sim$1629 MB & $\sim$195 GB \\
        water molecule (3 atoms)  & $\sim$1.5 MB  & $\sim$136 MB \\
        \bottomrule
    \end{tabular*}

    \vspace{2em}

    (b) Memory per MPI process --- slab+water (the bottleneck)\par\vspace{0.5em}
    \begin{tabular*}{\textwidth}{@{\extracolsep{\fill}}ccccc}
        \toprule
        \textbf{MPI Processes} & \textbf{Nodes (@ 4 MPI/node)} & \textbf{Peak / Process} & \textbf{Limit / Process} & \textbf{Fits?} \\
        \midrule
        32 & 8  & 6,935 MB & 6,656 MB & \ding{55} \\
        40 & 10 & 5,473 MB & 6,656 MB & \checkmark \\
        \bottomrule
    \end{tabular*}
\end{table*}

The agent reaches the dry run phase and then responds to the user with the following information:
\begin{lstlisting}
The slab calculations are large (~200 GB peak per process at 
hgrid=0.37). This is a 291-atom TiO2 slab - you will need a
significant HPC allocation with many MPI processes to make
the memory per process tractable.
\end{lstlisting}
The table of memory requirements generated in the LLMs response are listed in Table~\ref{tab:memory}. As the dry run tool allows the agent to specify the number of MPI ranks, we then ask the orchestrating agent:
\begin{lstlisting}
I plan to run this on supercomputer Fugaku. The calculation
should be run with 4 MPI per node. Each node can safely use
approximately 26 GB. Using dry run, tell me how many nodes I
should use for this calculation.
\end{lstlisting}
The agent performs two dry run tool calls to generate the required data. It then responds to the user:
\begin{lstlisting}
Recommendation: use 10 nodes (40 MPI processes, 4 per node).                                       
This gives ~1.2 GB of headroom per process on the worst-case
calculation (slab+water). If you want a more comfortable margin,
12 nodes (48 MPI) would bring the peak to ~4,560 MB/process,
leaving  ~2.1 GB free per process.
\end{lstlisting}
Thus, the dry-run approach provides a guard against the agent executing underprovisioned calculations. This saves the high latency work of finding a suitable calculation setup with each experiment going through the supercomputer queue and wasting computational resources.

% \subsection{The Need for Domain Specific Agentic Systems}

% Without adding any extra functionality, it is possible to perform these same sets of calculations with a sufficiently powerful AI agent. The agent could search through the publicly available source code of BigDFT to understand how a calculation is run. It could build a workflow and jobscript for running it. Finally, it could ssh into the supercomputer and submit the job. However, in order to reach the correct scientific result, an enormous amount of trial and error would be required; each submission to the supercomputer would require time waiting in the queue and use costly computer time. The design of \LARA is to ensure that this iterative process avoids the actual supercomputer until the last stages of the process to ensure timely scientific results.

\section{Discussion: Toward a Co-Piloted HPC Ecosystem}
\label{sec:discussion}

The results presented in this work highlight a key limitation of current AI-assisted scientific workflows: while modern large language models are highly effective at generating code, they lack the mechanisms required to ensure correctness and reproducibility in high-performance computing environments. By contrast, scientific simulations demand strict guarantees on input validity, resource usage, and physical consistency. In this work, we introduced \LARA, a validation-driven agentic framework designed to enable reliable generation of atomistic simulation workflows in HPC environments. By combining a controlled execution layer based on \RemoteManager with simulation-native validation through the dry-run capabilities of \BigDFT, we argued for a shift from generative AI workflows to a validation-first paradigm.

In principle, users may manage their atomistic simulations using any of the highly general agentic systems available today (coding agents, ``claws'', etc). These agents can run arbitrary commands, connect arbitrary systems, and generate arbitrary code. However, this extreme degree of non-deterministic behavior risks the wasting of precious computational resources on the production of scientifically meaningless results. When correct results are achieved, they will only come after long and costly trial-and-error loops. In fact, this iterate until correctness approach is the common strategy of current AI agents --- one that is effective only if the resource use, latency of trials, and safety risks are low.
Even when such a approach finally succeeds, human-in-the-loop would not be a warranted feature. Preserving traceable tool calling and well automated relationships between tool/phases of a production process will also ensure reproducibility of the scientific tasks.

For example, we noted in Sec.~\ref{subsec:operational} that \RemoteManager works by performing the same actions as a human user would; from this, we know that a general AI agent is also capable of connecting to a remote machine, creating and submitting jobscripts, checking for results, etc. For a given HPC resource or DFT program, when the underlying model is not sufficiently trained on the specific documentation, today's agents can search the web and gather the required information. Yet each of these setup steps delays achieving the goals of the scientific project. The benefit of providing suitable tooling is to be sure the agent does not need to waste time nor its context window on discovering the correct set of workflow operations.

Another weak point of deploying general agent systems is that each step in the research process becomes dependent on a frontier quality model. This may not be desireable due to the high costs and security risks that come with using commerical providers. Such an approach is also sensitive to changes to models made outside of the user's control, which may disrupt a given workflow. The modular LARA architecture separates stages by the quality of engine required (Fig.~\ref{fig:lara_architecture}), with tight inner reasoning loops accomplished using smaller models. The code that is generated becomes independent of any given model, allowing the workflow to be executed (even with a manual process) reproducibily in the future.

A simple solution to the weak points of general agents is to package predefined tools suitable for steps in a scientific workflow. For example, the composition of the model context protocol and \RemoteManager allows users to easily define tools that perform atomistic simulations using HPC resources. In addition to formalizing these steps to ensure immediate correctness and a separation of concerns, providing tools serves as a form of prompt engineering~\cite{10.1145/3560815} --- biasing agent actions towards approaches it may not immediately produce under standard operating conditions.

The drawback of this approach is that predefined tools are scientifically constraining. In research, the challenge is often not merely choosing among existing actions, but constructing a workflow variant that was not fully conceived in advance due to an evolving hypothesis. To allow for the full expressiveness of new procedural patterns, while ensuring controlled and correct execution, we introduce the \LARA AI system based on a domain specific, multi-agent loop. In addition to the dynamic prompting made possible by a multiphase workflow incorporating physics reasoning and RAG, \LARA is centered on a dry-run primitive that enables reliable validation of proposed workflows with low latency and resource use.

Based on these experiences, we argue that it is now the task of the broader computational physics community to develop shared knowledge on developing agentic systems suitable for its specific research requirements. These systems may come in the form of MCP servers, agent SKILL files, tools to be inserted in Agent Client Protocol schemes, or full packages like \LARA. The community's frameworks would then be integrated into end-to-end agentic systems like The AI Scientist~\cite{Lu2026-ks} and others~\cite{weng2026deepscientist,analemma2026fars,evoscientist2026,li2025rdagent,schmidgall2025agentlaboratoryusingllm,denario}, which are designed to autonomously conduct all distinct phases of scientific research --- research ideas, writes code, run experiments, plot and analyze data, write scientific manuscripts, and perform peer review. Recent research suggests that the length of tasks AI can reliably complete is doubling roughly every seven months~\cite{measuring-ai-ability-to-complete-long-tasks}. This implies that many of today's implementation and debugging limitations could be overcome sooner than expected. Nevertheless, integrating domain specific systems into these advanced models will remain essential to reduce the increasingly dominate burden of tool calling on agentic science~\cite{raj2025cpu}.

One such critical requirement will be
the ability to operate these systems within a \emph{sandboxed environment}. This environment must provide controlled access to simulation software and associated data, HPC resources and scheduling systems, external AI services or locally deployed models.
The sandbox ensures that all interactions are traceable, reproducible, and compliant with the constraints of the research context, including licensing restrictions and data confidentiality.
Within this framework, \LARA can be seen as an orchestration layer of the sandbox that connects state-of-the-art copilots, retrieval systems, and validation tools, without being tied to a specific provider or model. This flexibility is essential in scientific environments, where requirements vary widely across domains and institutions.

\vspace*{6pt}
\acknowledgements

W.D. acknowledges the FOCUS (Hyogo Prefecture, Kobe City) Project to Promote the Formation of a Research and Education Center (COE) in the Field of Computational Science. We acknowledge Eliott Jacopin for useful discussions.
We thank E. Polack, D. Caliste, J. Janssen, C. Herrera-Contreras, T. Abui Degbotse, J. Briki, L. Didukh and the organizers of the 2025 LLM Hackaton for Materials Science and Chemistry for stimulating discussions and for providing momentum to the initiative.

\bibliographystyle{apsrev4-2-titles}
\bibliography{references}
\end{document}